# Ionic-instability induced color tuning in lead-based, mixed-halide perovskites

Anthony Ruth[1,2,*], Halyna Okrepka[3], Michele Vergari[4], Charlie Desnoyers[3], Minh Nguyen[5], Luca Gavioli[4], Prashant V. Kamat[3,6], Masaru Kuno[2,3,6*]

[1] Chromatic Lighting, 6301 Gaston Ave STE 1321, Dallas, TX 75214 USA
[2] University of Notre Dame, Department of Physics and Astronomy, Notre Dame, IN 46556 USA
[3] University of Notre Dame, Department of Chemistry and Biochemistry, Notre Dame, IN 46556 USA
[4] I-LAMP and Dipartimento di Matematica e Fisica, Università Cattolica del Sacro Cuore, 48 Via della Garzetta, Brescia 25133, Italy
[5] PY Materials, 2 Davis Drive, Durham, NC 27709 USA
[6] Notre Dame Radiation Laboratory, Notre Dame, IN 46556 USA

aruth2@nd.edu
mkuno@nd.edu

**Introduction**

Mixed-halide perovskites such as methylammonium lead iodide-bromide [$MAPb(I_{1-x}Br_x)_3$], represent a new class of ionically-active semiconductors whose band gaps ($E_g$) can be tuned via halide stoichiometry ($x$). These materials have been investigated for use in applications ranging from tandem solar cells[1] to light emitters.[2] However, when illuminated with continuous wave (CW) excitation intensities ($I_{exc}$) close to 1 sun, their band gaps change. This behavior is most readily observed as redshifts of the mixed-halide perovskite photoluminescence (PL) spectrum under illumination. Observed redshifts are linked through direct structural measurements to the nucleation and growth of low band gap, halide-enriched inclusions within parent, mixed-halide materials. For $MAPb(I_{1-x}Br_x)_3$, this reflects the nucleation/growth of iodine-rich (I-rich) domains.[3] Under darkness, the PL gradually blueshifts back to that of the original parent phase. Structural measurements confirm that photosegregation reverses itself with entropically-driven anion remixing restoring the material to its initial, mixed state.[3]

Contrary to conventional wisdom, intermediate photoluminescence energies can be stabilized in mixed-halide perovskites during photosegregation. It is commonly assumed that photosegregation near universally leads to fixed terminal energies consistent with a final stoichiometry of $x_{terminal}$~0.2 [i.e., $E_{terminal,x=0.2}$].[3] However, we show that intermediate colors/stoichiometries, between those of the parent mixed-halide alloy and $E_{terminal,x=0.2}$, can be stabilized by varying pulsed laser excitation repetition rates and peak fluences. This demonstrates rudimentary color tuning and points to an opportunity to capitalize on mixed-halide perovskite ionic photoinstabilities to generate tunable colors on demand. At a more fundamental level, such color tuning begs the question of how intermediate photosegregation energies arise and how they are kinetically stabilized given conflicting reports in the literature regarding photosegregation's pulsed excitation repetition rate and peak fluence dependencies.

In this regard, much effort has been devoted to establishing a microscopic mechanism for mixed-halide perovskite photosegregation. Posited models include thermodynamic[4,5,6,7,8,9,10] chemical,[11,12] polaron,[13,14,15] and trap-related explanations.[16,17,18] Models vary in their degree of quantitativeness with the majority being descriptive. The band gap thermodynamic model stands as an exception, predicting photosegregation excitation intensity thresholds,[4,7,9] temperature-independent $x_{terminal}$-values,[19] and composition-dependent, photosegregation proclivities.[9]

More importantly, the band gap thermodynamic model predicts that intermediate terminal energies and stoichiometries ($E_{terminal}$ and $x_{terminal}$) can be stabilized.[4,5,6] Within the model, intermediate $E_{terminal}$ and $x_{terminal}$ are a consequence of carrier density ($n$)-dependent suppression of entropic contributions to mixing free energies ($\Delta F^*$). Reference 4 provides more details about the band gap thermodynamic model.

Irrespective of model, limited work has focused on developing a practical, kinetic understanding of how intermediate $E_{terminal}$ and $x_{terminal}$ values can be realized, let alone how color tuning can be achieved under pulsed laser excitation. The little kinetic information that is known point to photosegregation forward rate constants ($k_{forward}$) being $I_{exc}$- and $T$-dependent. $k_{forward}$-values are additionally up to four orders of magnitude larger than corresponding, dark remixing rate constants ($k_{reverse}$).[9,20,21,22]

We now demonstrate that intermediate $E_{terminal}$ ($x_{terminal}$)-values can be obtained through pulsed laser excitation of formamidinium/cesium lead iodide/bromide [$FACsPb(I_{1-x}Br_x)_3$] thin films. We further develop a kinetic rationalization for intermediate color stabilization through concerted CW and pulsed laser photosegregation and dark remixing measurements. Experiments are accompanied by Kinetic Monte Carlo (KMC) simulations. What emerges is a practical and self-consistent explanation of mixed-halide perovskite photosegregation/dark remixing kinetics with implications for mixed-halide perovskite lighting applications. This kinetics-based understanding of photosegregation may ultimately help explain other reported (and less understood) photosegregation phenomena such as pulsed laser-induced photoremixing and spectral blueshifting.[23,24] Details of both the optical measurements and KMC simulations have been provided in the article's Supplementary Information (SI) as well as in **Figures S1-S4**.

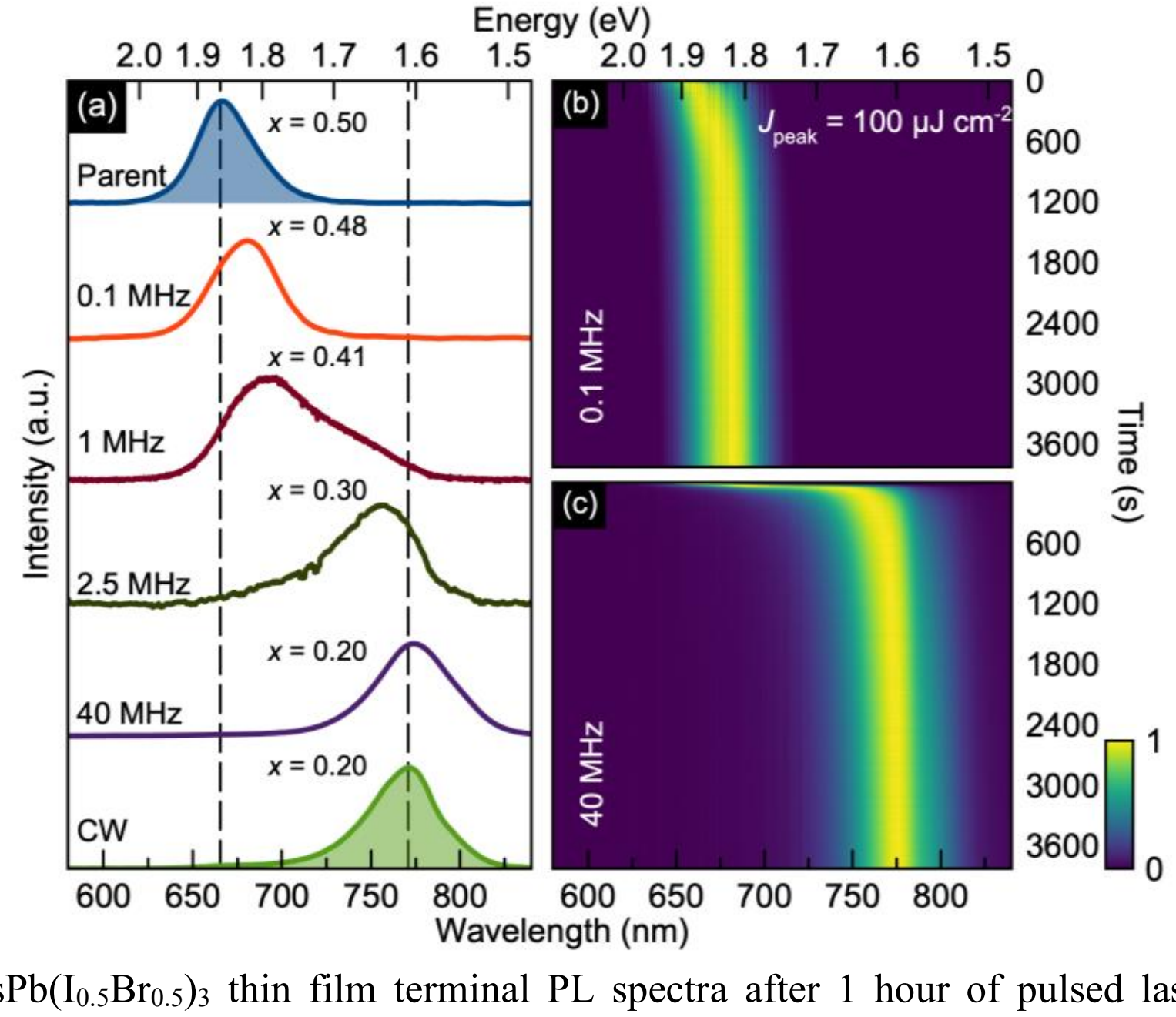

**Figure 1**. (a) $FACsPb(I_{0.5}Br_{0.5})_3$ thin film terminal PL spectra after 1 hour of pulsed laser excitation. Laser repetition rates varied under a constant peak fluence of $J_{peak}$=100 µJ cm$^{-2}$. Parent, mixed-halide and fully CW segregated ($I_{exc}$=3 W cm$^{-2}$) PL spectra shown for comparison purposes. Vertical dashed lines correspond to their PL energies. (b) Corresponding heatmaps for 0.1 MHz and (c) 40 MHz photosegregation spectral time series with spectra normalized to one.

**Results and Discussion**

**Figure 1a** first illustrates intermediate $E_{terminal}$ ($x_{terminal}$) observed in $FACsPb(I_{0.5}Br_{0.5})_3$ ($x$=0.5) thin film spectra under pulsed laser excitation (constant peak fluence of $J_{peak}$=100 μJ cm$^{-2}$, variable repetition rates, $f$, between 0.1-40 MHz). For comparison purposes, emission spectra of the parent alloy and a fully (CW) photosegregated film with $x_{terminal}$~0.2 ($E_{terminal,x=0.2}$=1.67 eV) are shown. Observed $E_{terminal}$ have been converted to corresponding $x_{terminal}$-values using an empirical Vegard's law relationship for $FACsPb(I_{1-x}Br_x)_3$.[6,25,26] Evident are stabilized, intermediate $E_{terminal}$ ($x_{terminal}$) values, achieved using pulsed laser excitation.

**Figures 1b** and **1c** illustrate spectral trajectories under reported pulsed excitation conditions over the course of 1 hour. For all traces, once the peak emission energy reaches a given $E_{terminal}$, it remains stable without any further redshifts during the measurement. **Figure S5** summarizes analogous data acquired using a peak fluence of $J_{peak}$=10 μJ cm$^{-2}$. For the current study, all pulsed laser photosegregation spectra revert back to that of the parent alloy over the course of 24 hours. A representative dark recovery spectrum is provided in SI (**Figure S6**).

To understand above pulsed laser $E_{terminal}$ dependencies, we return to more established mixed-halide perovskite photosegregation and dark remixing kinetics under CW excitation conditions. Here, reports now establish order of magnitude, room temperature photosegregation $k_{forward}$-values between $10^{-3}$ and 1.0 s$^{-1}$. [9,20,21,22] An exact estimate of $k_{forward}$ is complicated by $I_{exc}$-[9], initial stoichiometry ($x_{init}$)[3], and temperature-dependencies[22], the latter showing Arrhenius-like behavior.[22] Additional complications come from environmental[16], surface passivation[17], and possible film quality/thickness dependencies.[18,27]

Complementary $k_{reverse}$-values range from $k_{reverse}$~$3.8\times10^{-4}$-$6.8\times10^{-3}$ s$^{-1}$ and are up to four orders of magnitude smaller than $k_{forward}$.[9,20,21,22] This results in associated dark remixing timescales on the order of hours, in contrast to photosegregation, which occurs over minutes.[3,9] $k_{reverse}$ is $T$-dependent and, like $k_{forward}$, increases with increasing $T$ in an Arrhenius fashion.[22] A corresponding $E_a$-value of 0.55 eV (53.5 kJ mol$^{-1}$)[22] aligns with previously reported anion migration activation energies of 0.09-0.27 eV (8.7-26.1 kJ mol$^{-1}$) for $Br^-$ and 0.08-0.58 eV (7.7-56.0 kJ mol$^{-1}$) for $I^-$.[28,29,30,31,32] This supports the conclusion that photosegregation and dark remixing involve vacancy-mediated anion migration. It is further corroborated by direct experimental and KMC studies.[7,33] **Table S1** of the SI tabulates all literature-reported $k_{forward}$- and $k_{reverse}$-values.

**Figure 2a** now shows representative, room temperature CW photosegregation and dark remixing spectra ($I_{exc}$=0.5 W cm$^{-2}$). Spectra have been acquired over the course of 1 hour of photosegregation and 17 hours of dark remixing. Analogous measurements have been conducted for different $I_{exc}$ (**Figure S7**).

In all cases, average PL energies, $\langle E_{PL}\rangle$, of redshifted (i.e., photosegregated) contributions to the total emission have been extracted using Gaussian fits to better establish photosegregation and dark remixing kinetics. The SI and **Figure S8** provide details of the $\langle E_{PL}\rangle$ fitting/extraction procedure.

**Figures 2c** and **2d** show resulting $FACsPb(I_{0.5}Br_{0.5})_3$ photosegregation and dark recovery $\langle E_{PL}\rangle$ trajectories (denoted $\langle E_{light}\rangle$ and $\langle E_{dark}\rangle$ respectively in what follows) for CW photosegregation $I_{exc}$-values of $I_{exc}$=0.025, 0.5 and 3.0 W cm$^{-2}$. **Figure S9** shows analogous trajectories for photosegregation $I_{exc}$-values of $I_{exc}$=0.05, 0.1, and 1.0 W cm$^{-2}$. In whole, the data reveal that $\langle E_{\text{light}} \rangle$ trajectories exhibit sigmoidal decays with an induction period ($\tau_0$) prior to photosegregation.

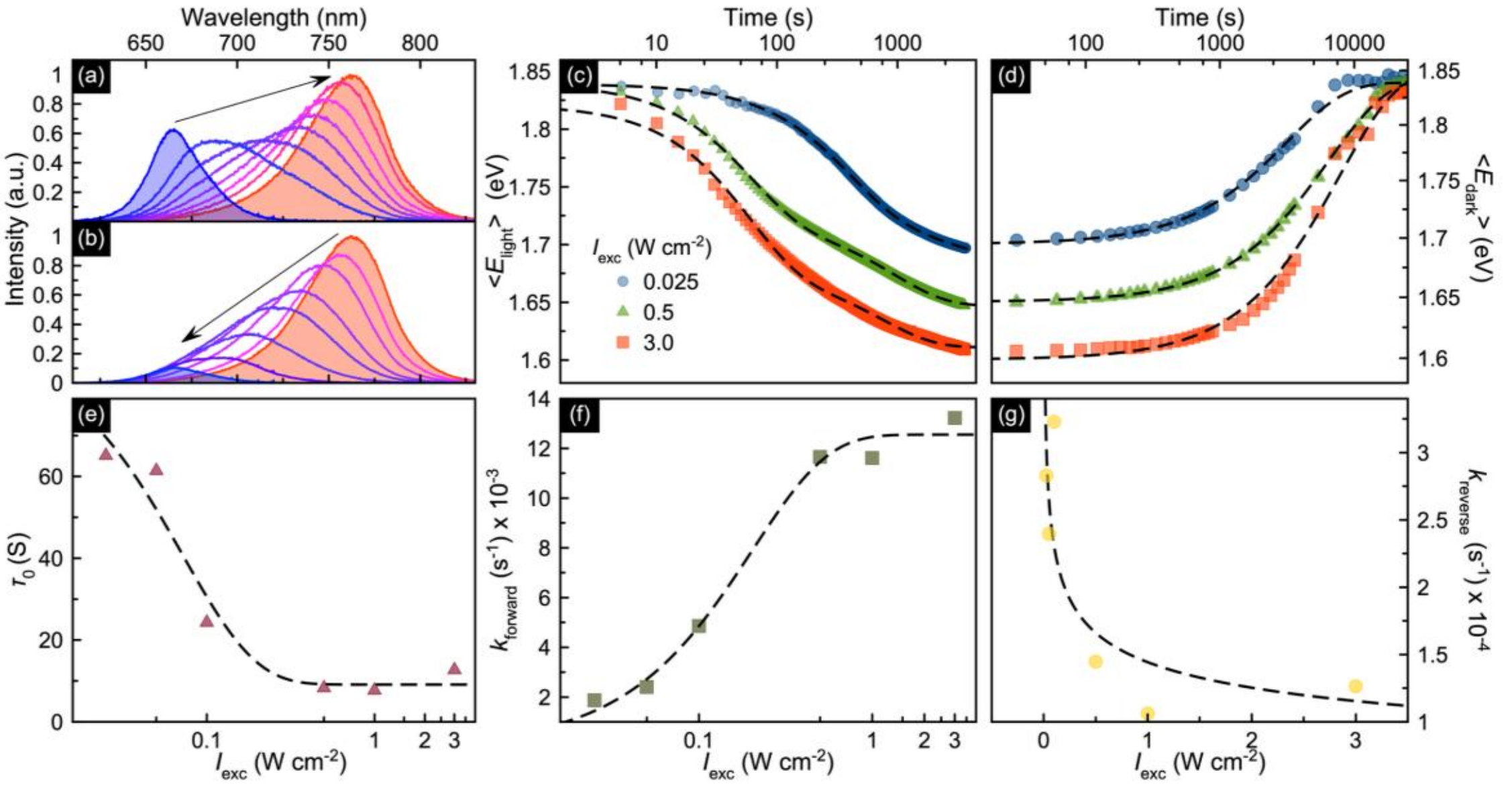

**Figure 2**. $FACsPb(I_{0.5}Br_{0.5})_3$ room temperature CW (a) photosegregation and (b) dark remixing spectra for $I_{exc}$=0.5 W cm$^{-2}$. In (a) initial ($t$=0 s) and final ($t$=3600 s) spectra shown in blue and red, respectively; Intermediate spectra extracted at $t$=50, 100, 200, 400, 800 and 1600 s. In (b) initial ($t$=0 s) and final ($t$=61,200 s) spectra shown in red and blue, respectively; Intermediate spectra extracted at $t$=500, 1500, 2500, 4000, 5000 and 9000 s. $I_{exc}$-dependent (c) $\langle E_{\text{light}} \rangle$ and (d) $\langle E_{\text{dark}} \rangle$ trajectories for $I_{exc}$=3.0 (red), 0.5 (green) and 0.025 (blue) W cm$^{-2}$. Corresponding fits shown using dashed lines. Experimental $I_{exc}$ -dependence of (e) $\tau_0$ (dashed line, guide to the eye), (f) $k_{forward}$ (dashed line, fit to **Equation 2**), and (g) $k_{reverse}$ (dashed line, guide to the eye).

$\langle E_{\text{light}} \rangle$ trajectories have consequently been fit to a time-delayed, sigmoid of the form

$$\langle E_{\text{light}}(t) \rangle = E_{\text{init}} - \Delta E\left[1 - e^{-k_{\text{forward}}(t-\tau_0)}\right] \tag{1}$$

to extract estimates of $\tau_0$ and $k_{forward}$. In **Equation 1**, $E_{init}$ is the initial PL energy of the mixed-halide parent phase and $\Delta E = E_{\text{init}} - \langle E_{\text{terminal}} \rangle$ is the energy difference between the initial, mixed-halide emission energy and that of the final, segregated PL. Details of the fitting procedure, including summary of extracted parameters, have been provided in the SI (**Table S2**, **Figure S10**).

Fits are shown as dashed lines in **Figure 2c**. They highlight how **Equation 1** captures both the induction time and overall S-like evolution of $\langle E_{\text{light}}(t) \rangle$. **Figure 2e** plots extracted $\tau_0$, revealing that it decreases 5-fold with increasing $I_{exc}$ ($\tau_0$~65 s, $I_{exc}$=0.025 W cm$^{-2}$; $\tau_0$~13 s, $I_{exc}$=3.0 W cm$^{-2}$). $\tau_0$ has previously been noted by Herz whose reported $\tau_0$-values range from 100 s for $I_{exc}$=0.07 W cm$^{-2}$ to 4 s for $I_{exc}$=300 W cm$^{-2}$, both at 295 K.[34] As will be seen below, the induction time originates from the slow growth of nascent I-rich domains.

**Figure 2f** shows accompanying fit-extracted $k_{forward}$-values, which grow exponentially with $I_{exc}$ [or corresponding carrier density, $n = \frac{\alpha I_{\text{exc}}}{h\upsilon}\tau$; $\alpha = 3.1 \times 10^5$ cm$^{-1}$ is the absorption coefficient (**Figure S11**), $h\upsilon$ is the photon energy, and $\tau$=2.6 ns is an effective carrier lifetime (**Figure S12**)] until saturation. Extracted $k_{forward}$-values are fit to

$$k_{\text{forward}}(I_{\text{exc}}) = k_{\text{sat}}(1 - \mathrm{e}^{-\beta n}) \tag{2}$$

with $k_{\text{sat}} = 0.012$ $\mathrm{s}^{-1}$ an empirical saturation rate constant and $\beta = 1.5 \times 10^{-15}$ $\mathrm{cm}^3$ a fit parameter. **Equation 2**'s functional form is motivated by prior observations and rationalizations of $k_{\text{forward}}$ exponential growth and saturation in $MAPb(I_{1-x}Br_x)_3$ thin films.[9,34]

**Figure 2d** plots complementary $\langle E_{\text{dark}} \rangle$ trajectories, which do not exhibit an induction time and which are well-represented by the following exponential function,

$$\langle E_{\text{dark}}(t) \rangle = E_{\text{init}} - \Delta E e^{-k_{\text{reverse}} t}. \tag{3}$$

Dashed lines in **Figure 2d** are fits to experiment using the above expression.

**Figure 2g** shows fit-extracted $k_{\text{reverse}}$-values. Of note are $k_{\text{reverse}}$-values that decrease with increasing photosegregation $I_{\text{exc}}$. This photosegregation $I_{\text{exc}}$-dependent $k_{\text{reverse}}$ behavior is not captured by early, empirical $k_{\text{reverse}}$ expressions of the form $k_{\text{reverse}} = Ae^{-\frac{E_a}{RT}}$.[22,25]

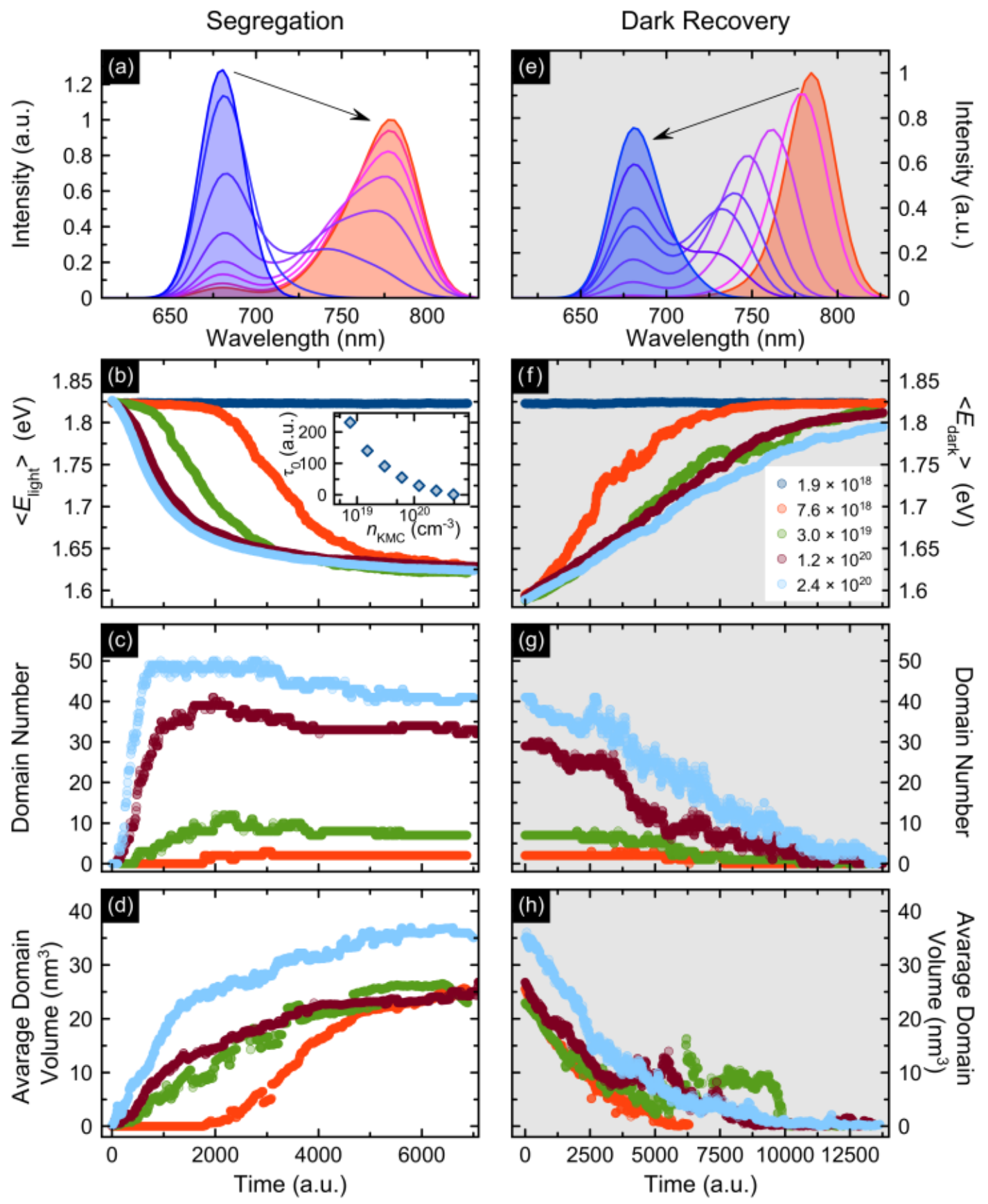

**Figure 3**. KMC photosegregation (a) PL spectra, (b) <$E_{\text{light}}$> trajectories, Inset: $\tau_0$, (c) domain number, and (d) average domain volume. KMC dark recovery (e) PL spectra, (f) <$E_{\text{dark}}$> trajectories, (g) domain number, and (h) average domain volume. For (a,e), $n_{\text{KMC}}$=3.0×10$^{19}$ cm$^{-3}$. For (b,f) $n_{\text{KMC}}$=1.9×10$^{18}$-2.4×10$^{20}$ cm$^{-3}$ [1.9×10$^{19}$ (navy blue), 7.6×10$^{18}$ (red), 3.0×10$^{19}$ (green), 1.2×10$^{20}$ (dark red), 2.4×10$^{20}$ (light blue)]. Associated (e,f) domain number during (e) photosegregation and (f) dark recovery and (g,h) average domain volume during (g) photosegregation and (h) dark recovery.

To address the microscopic origin of $\tau_0$ and above $k_{\text{reverse}}$ $I_{\text{exc}}$ dependencies, we conduct CW KMC simulations. **Figure 3a** shows KMC-derived photosegregation PL spectra for given carrier concentration ($n_{KMC}$=3.0×10$^{19}$ cm$^{-3}$). Simulated spectra redshift with increasing KMC steps and closely reproduce the experiment in **Figure 2a**. **Figure 3b** summarizes associated KMC-derived <$E_{light}$> trajectories for $n_{KMC}$=1.9×10$^{18}$-2.4×10$^{20}$ cm$^{-3}$. With the exception of $n_{KMC}$=1.9×10$^{18}$ cm$^{-3}$, all <$E_{light}$>-trajectories exhibit sigmoidal behavior and converge to $\langle E_{\text{terminal},x=0.2}\rangle$=1.626 eV. That a critical $n_{KMC}$ exists for <$E_{light}$> to decrease agrees with earlier, empirically-established $I_{exc,threshold}$-values for photosegregation.[4,7] More notable though are KMC <$E_{light}$> trajectories that exhibit induction periods prior to photosegregation. Estimated KMC induction times decrease by 20 with increasing $n_{KMC}$ between 7.6×10$^{18}$ and 2.4×10$^{20}$ cm$^{-3}$.

From acquired KMC snapshots, we obtain microscopic insight into $\tau_0$. **Figure 3c** shows that the number of nucleated I-rich domains increases from 3 for $n_{KMC}$=7.6×10$^{18}$ cm$^{-3}$ to 49 for $n_{KMC}$=2.4×10$^{20}$ cm$^{-3}$. Nucleated domains are small and have associated volumes that span a relatively narrow range between 20-40 nm$^3$ (effective radii, $r$=1.7-2.1 nm). At early times, domains grow slowly and exhibit little to no spectral redshifting. This stems from nucleated domains possessing local band gaps that make carrier funneling to them inefficient. **Figures 3c** and **3d** summarize the evolution of KMC domain number and average volume during photosegregation.

**Figures 3e** and **3f** plot associated KMC dark remixing spectra and <$E_{dark}$> trajectories. Both reveal entropically-driven recovery back to the parent alloy. KMC $k_{reverse}$-values are nine-fold smaller than KMC $k_{forward}$-values (**Figure S13**) and capture experimentally-observed asymmetries in forward and reverse photosegregation/dark remixing kinetics. More interesting are corresponding changes to underlying domain numbers (**Figure 3g**) and average volumes (**Figure 3h**). We find that domain volumes decay at a similar rate irrespective of their initial (terminal) sizes following photosegregation. Domain numbers, by contrast, decrease more slowly with increasing domain number at the outset of dark remixing. We conclude that the persistence of a few domains from a large, initial number is why experimental $k_{reverse}$-values decrease with increasing photosegregation $I_{exc}$.

Above CW experimental and KMC kinetic analyses now inform our understanding of pulsed laser photosegregation and dark remixing behavior. Under pulsed laser excitation, photogenerated carrier densities, $n$, in the mixed-halide perovskite are not constant as in the CW (steady state) case. Rather, they are time-dependent and cyclical. For convenience, we assume exponential decays in what follows, i.e., $n(t) = n_{\text{peak}}e^{-\frac{t}{\tau}}$ with $n_{\text{peak}} = \frac{J_{\text{peak}}\alpha}{h\nu}$ a peak photogenerated carrier density. Other decay profiles do not change the overall predicted behavior. An average $k_{forward}$ during pulsed illumination is therefore

$$\langle k_{\text{forward}}\rangle = \frac{1}{\tau_{\text{cycle}}}\int_0^{\tau_{\text{cycle}}} k_{\text{sat}}\left(1 - e^{-\beta n(t)}\right) dt \qquad (4)$$

with $\tau_{\text{cycle}} = \frac{1}{f}$ the period and $k_{\text{sat}}$ and $\beta$ from the $k_{forward}$ fit in **Figure 2f**.

**Figure 4a** now shows terminal, pulsed excitation (constant peak fluence, $J_{peak}$=100 μJ cm$^{-2}$; $n_{\text{peak}} = 2 \times 10^{19}$ cm$^{-3}$) photosegregation spectra for progressively larger pulsed laser repetition rates between $f$=0.1-40 MHz ($\tau_{cycle}$=10-0.025 μs). **Figure S14** shows the full spectral data set over 1 hour. **Figure S15** provides analogous 1 hour data for measurements conducted with constant $J_{peak}$=10 μJ cm$^{-2}$.

Experimental $\langle k_{\text{forward}}\rangle$-values are extracted by fitting individual $\langle E_{\text{light}}(t)\rangle$ trajectories, underlying the terminal spectra in **Figure 4a** and **Figure S5**, with **Equation 1**. $\langle E_{\text{light}}(t)\rangle$ trajectories and fits for $J_{peak}$=100 μJ cm$^{-2}$ and 10 μJ cm$^{-2}$ have been provided in the SI (**Figure**

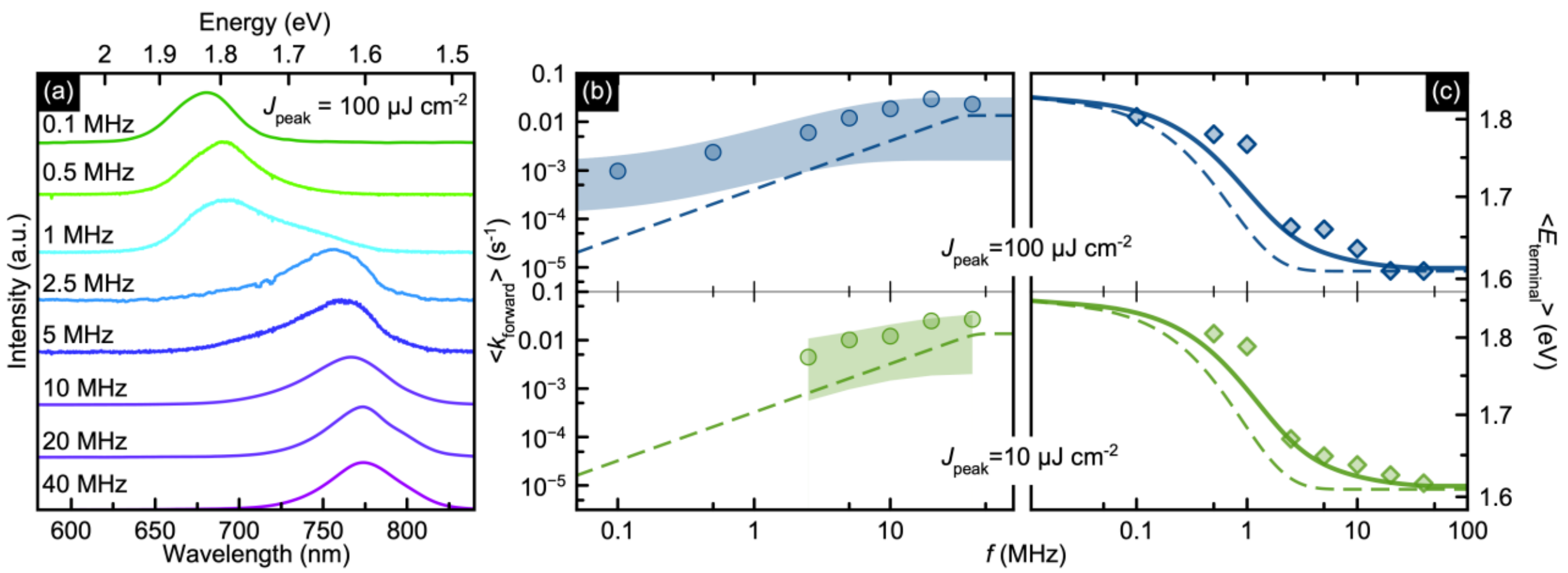

**Figure 4**. (a) $FACsPb(I_{0.5}Br_{0.5})_3$ thin film terminal PL spectra after 1 hour of pulsed laser excitation. Excitation conditions: $J_{peak}$=100 μJ cm$^{-2}$; $f$=0.1-40 MHz. (b) Experimental $<k_{forward}>$ for $f$=0.1-40 MHz with $J_{peak}$=100 μJ cm$^{-2}$ (top) and 10 μJ cm$^{-2}$ (bottom). Dashed lines: $<k_{forward}>$ predictions from **Equation 4**. Shaded regions: area defined by upper and lower $k$-values in fits to data. (c) $\langle E_{terminal}\rangle$ trajectories for $f$=0.1-40 MHz with $J_{peak}$=100 μJ cm$^{-2}$ (top) and 10 μJ cm$^{-2}$ (bottom). Lines: predictions from **Equations 4**/**1** (dashed lines) and **Equation 10** (solid lines).

**S16**). **Figure 4b** plots obtained $\langle k_{forward}\rangle$-values and reveals that they increase with increasing $f$ (decreasing $T$) in a near linear fashion. Shaded regions denote an extracted $\langle k_{forward}\rangle$ range, based on upper and lower $k$-values obtained in fits to the data. The SI provides more details of the fitting procedure as well as a semiquantitative rationalization of $\langle k_{forward}\rangle$'s near linear $f$ dependence.

Plotted atop experimental $\langle k_{forward}\rangle$-values are predictions from **Equation 4** for $J_{peak}$=100 μJ cm$^{-2}$ and 10 μJ cm$^{-2}$ (dashed lines). The general fidelity of **Equation 4** to experimental $\langle k_{forward}\rangle$-values highlights how pulsed excitation measurements are essentially abbreviated versions of the CW experiment provided that $k_{reverse}t \ll 1$ (threshold at $t$~10$^2$-10$^4$ s for $k_{reverse}$=10$^{-4}$-10$^{-2}$ s$^{-1}$).

**Figure 4c** now summarizes extracted (intermediate) $\langle E_{terminal}\rangle$ from **Figure 4a** as a function of $f$. Evident is that $\langle E_{terminal}\rangle$ decreases with increasing $f$ (increasing excitation duty cycle, $\eta$) in a sigmoidal fashion. Also shown is the effect of varying $J_{peak}$ wherein reducing it from $J_{peak}$=100 μJ cm$^{-2}$ to 10 μJ cm$^{-2}$ causes $\langle E_{terminal}\rangle$ $f$-dependences to shift towards higher $f$. Plotted atop the experimental data are corresponding $\langle E_{terminal}\rangle$, predicted by **Equation 4** used in conjunction with **Equation 1** (dashed lines). There is again general agreement between the two with the simulated data capturing the overall shape and magnitude of $\langle E_{terminal}\rangle$. Solid lines show $\langle E_{terminal}\rangle$ when accounting for the reverse process as described below.

Given that **Equation 4** best applies in the limit where $k_{reverse}t \ll 1$ (i.e., dark, entropic remixing need not be considered), we conduct the following kinetic analysis to fully illustrate the influence of $k_{reverse}$ on pulsed segregation kinetics. In what follows, $\langle E_{PL}\rangle$ is treated as a memoryless state function. Under pulsed illumination, the mixed-halide perovskite advances along the CW illumination trajectory (**Equation 1**). An average photosegregation rate constant, $\langle k_{forward}\rangle$, characterizes the photosegregation kinetics over a full cycle. Between pulses, entropically-driven dark remixing advances the system along a CW dark trajectory (**Equation 3**), defined by the laser repetition rate. The change in $\langle E_{PL}\rangle$ during the illumination pulse is then dictated by a forward photosegregation contribution

$$\frac{d\langle E_{light}\rangle}{dt} = \langle k_{forward}\rangle(E_{init} - \Delta E - \langle E_{light}\rangle), \tag{5}$$

accompanied by a corresponding, reverse change during the ensuing dark period

$$\frac{d\langle E_{\mathrm{dark}}\rangle}{dt} = k_{\mathrm{reverse}}(E_{\mathrm{init}} - \langle E_{\mathrm{dark}}\rangle). \tag{6}$$

Implicit in **Equation 6** is the assumption that the dark period is near equivalent to $\tau_{\mathrm{cycle}}$.

During pulsed illumination, a dynamic equilibrium establishes itself between forward photosegregation and reverse, dark remixing with the following conditions

$$\langle E_{\mathrm{light}}\rangle = \langle E_{\mathrm{dark}}\rangle \tag{7}$$

and

$$\frac{d\langle E_{\mathrm{PL}}\rangle}{dt} = \frac{d\langle E_{\mathrm{light}}\rangle}{dt} + \frac{d\langle E_{\mathrm{dark}}\rangle}{dt}. \tag{8}$$

From **Equations 5-8**, it can be shown that

$$\frac{d\langle E_{\mathrm{PL}}\rangle}{dt} = -k_{\mathrm{net}}\langle E_{\mathrm{PL}}\rangle + [k_{\mathrm{net}}E_{\mathrm{init}} - \langle k_{\mathrm{forward}}\rangle\Delta E] \tag{9}$$

with $k_{\mathrm{net}} = \langle k_{\mathrm{forward}}\rangle + k_{\mathrm{reverse}}$. **Equation 9** has been derived in the SI. Solving for $\langle E_{\mathrm{PL}}\rangle$ in **Equation 9** then gives

$$\langle E_{\mathrm{PL}}(t)\rangle = E_{\mathrm{init}} - \Delta E\,\frac{\langle k_{\mathrm{forward}}\rangle}{k_{\mathrm{net}}}(1 - e^{-k_{\mathrm{net}}t}) \tag{10}$$

and captures the sigmoidal form of $\langle E_{\mathrm{light}}(t)\rangle$, first seen in **Equation 1**. **Figure S17** demonstrates that for short times $\langle E_{\mathrm{PL}}(t)\rangle$ can be understood using a $k_{\mathrm{forward}}$ only analysis (i.e., **Equations 4** and **1**). Longer experiments, however, require inclusion of $k_{\mathrm{reverse}}$ via **Equation 10**.

At equilibrium where changes in $\langle E_{\mathrm{PL}}\rangle$ due to photosegregation exactly match that due to dark remixing

$$\langle E_{\mathrm{terminal}}\rangle = \langle E_{\mathrm{PL}}(t \to \infty)\rangle = E_{\mathrm{init}} - \Delta E\,\frac{\langle k_{\mathrm{forward}}\rangle}{\langle k_{\mathrm{forward}}\rangle + k_{\mathrm{reverse}}} \tag{11}$$

**Equation 11's** $f$-dependence better reveals itself when a square pulse excitation profile is assumed. In this case, $\langle k_{\mathrm{forward}}\rangle$ can be approximated as $\langle k_{\mathrm{forward}}\rangle = \eta_{\mathrm{eff}}k_{\mathrm{forward}}$ where $\eta_{\mathrm{eff}}$, as described in the SI, is an effective excitation duty cycle, proportional to $f$. In this case, **Equation 11** reduces to the simpler form:

$$\langle E_{\mathrm{terminal}}\rangle = E_{\mathrm{init}} - \Delta E\,\frac{\eta_{\mathrm{eff}}}{\eta_{\mathrm{eff}} + \left(\frac{k_{\mathrm{reverse}}}{k_{\mathrm{forward}}}\right)} \tag{12}$$

For large duty cycles (large $f$), where $\eta_{\mathrm{eff}} \gg \frac{k_{\mathrm{reverse}}}{k_{\mathrm{forward}}}$ (alternatively, $\eta_{\mathrm{eff}} = 1$ in the CW limit), $\langle E_{\mathrm{terminal}}\rangle \to E_{\mathrm{terminal},x=0.2}$. For small duty cycles (small $f$) where $\eta_{\mathrm{eff}} \to 0$, $\langle E_{\mathrm{terminal}}\rangle \to E_{\mathrm{init}}$. This immediately rationalizes prior observations of near complete photosegregation for large $f$ and, conversely, suppressed photosegregation at low $f$.[18,35] The model also explains, via **Equation 4**, the near-linear increase in $\langle k_{\mathrm{forward}} \rangle$ with $f$ under constant $J_{\mathrm{peak}}$, as reported by Koc *et al*.[36]

The model and **Equation 12** rationalize other seemingly contradictory literature data. Specifically, Knight *et al*.[16] conduct variable repetition rate photosegregation measurements with $\eta$=0.5. They report no $f$ dependencies for $f$=10-3500 Hz and consequently propose that the fraction of trap-mediated recombination, experienced by a specimen under a given excitation condition, dictates photosegregation. While possible, we note that **Equation 12** immediately explains $f$ insensitive $\langle E_{\mathrm{PL}} \rangle$ in their measurements. This is because, as conducted, illumination pulses possess a constant $I_{\mathrm{exc}}$, making $k_{\mathrm{forward}}$ a constant. Additionally, $\eta$ is deliberately made $f$ independent due to changes of the square wave temporal width when altering repetition rates. Predicted are therefore near identical photosegregation kinetics (**Equation 2**) and terminal energies (**Equation 12**) between measurements. **Figure S18** repeats the Knight experiment via KMC simulation, holding $\eta$ fixed at $\eta$=0.2 and varying $f$ between $f$=5×10$^{-7}$ and $f$=10$^{-3}$ step$^{-1}$. It reveals nearly $f$-independent $\langle E_{\mathrm{PL}} \rangle$ trajectories in agreement with both the experiment and the developed model.

We now invoke KMC simulations to establish deeper insight into pulsed laser photosegregation. **Figures 5a** and **5b** plot KMC-derived $\langle E_{\mathrm{light}} \rangle$ for square wave and decaying square wave excitation profiles. The decaying square wave profile models persistent carriers in the material, following photogeneration. For either, $\langle E_{\mathrm{light}} \rangle$ decays with increasing frequency (duty cycle) in agreement with **Figure S16**. Occasional $\langle E_{\mathrm{light}} \rangle$ drift occurs for intermediate $f$ in the decaying square wave case. The most important results to emerge, though, are KMC-predicted intermediate $E_{terminal}$, which corroborate **Figures 1** and **4c**.

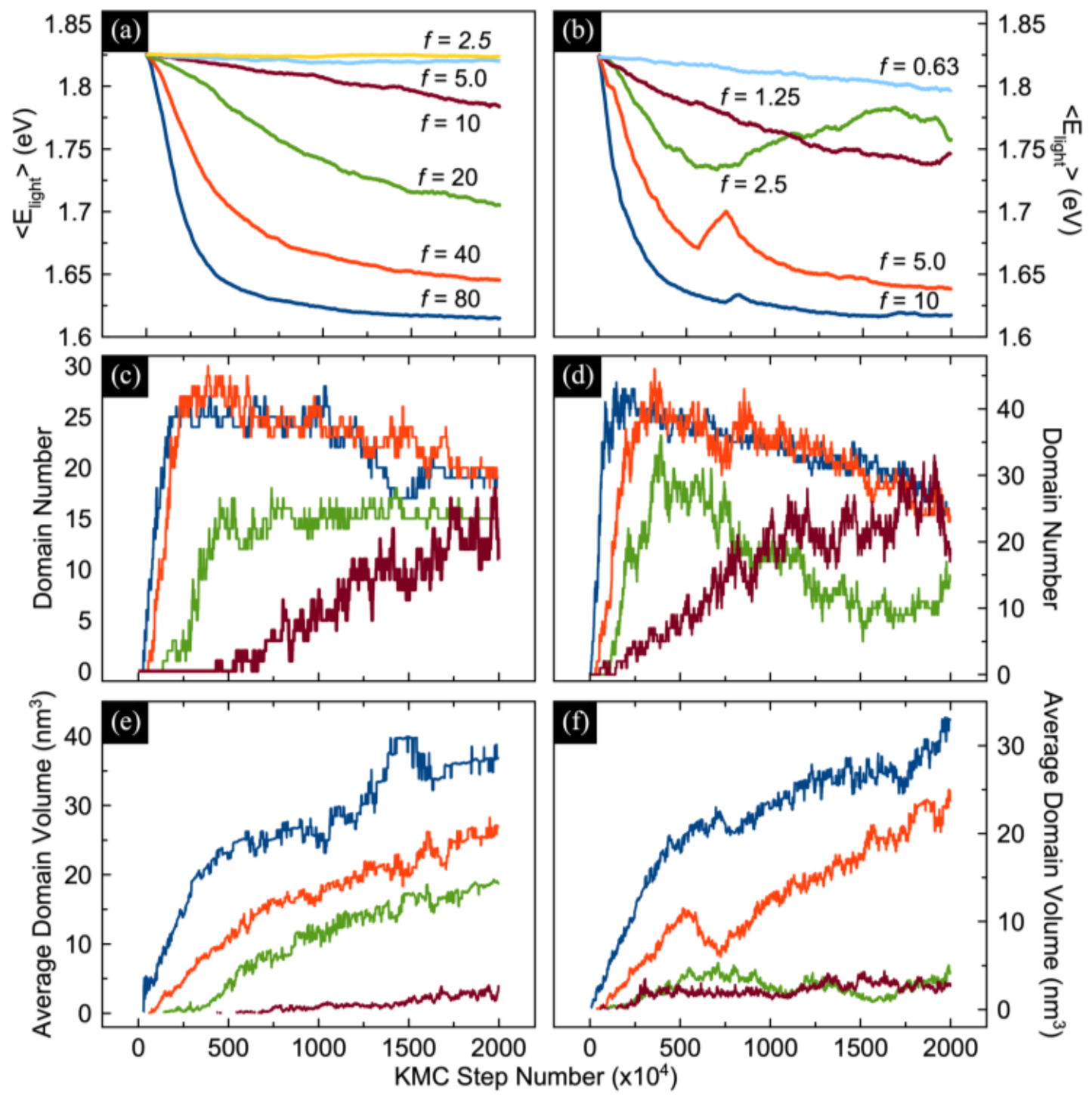

**Figure 5**. Variable photosegregation $f$ ($\eta$)-dependent KMC (a,b) $\langle E_{\mathrm{light}} \rangle$ trajectories for (a) square wave and (b) decaying square wave excitation profiles, (c,d) domain numbers for (c) square wave and (d) decaying square wave profiles, (e,f) average domain volumes for (e) square wave and (f) decaying square wave profiles. Frequencies are in units of 10$^{-4}$ step$^{-1}$.

These KMC simulations also reveal differences in photosegregation behavior between pulsed and CW excitation conditions. Specifically, we observe terminal, photosegregated domain sizes to increase with increasing *f*. For square wave [decaying square wave] excitation profiles with repetition rates between $f$=$10^{-3}$-$10^{-2}$ step$^{-1}$ ($\eta$=0.10-0.80) [$f$=$10^{-4}$-$10^{-3}$ step$^{-1}$ ($\eta$=0.01-0.10)], terminal volumes increase from 3 to 37 nm$^3$ [2.6 to 32 nm$^3$]. This is unlike the CW case where terminal volumes remain relatively similar between $n_{KMC}$.

A final distinction between pulsed and CW excitation results is the observation of Ostwald ripening, which supplants induction behavior. This is illustrated in **Figures 5c** and **5d**, which show square wave and decaying square wave domain numbers initially rising but then eventually falling over time. The decay is most acute for the $f$=$2.5\times10^{-4}$ step$^{-1}$ ($\eta$=0.025) decaying square wave trajectory where the domain number decreases from 35 to 10 by the end of the simulation. **Figures 5e** and **5f** point to the growth of large domains at the expense of smaller ones. Observed Ostwald ripening is responsible for the $<E_{light}>$ drift in **Figure 5b** and consequently stands as a potential instability for achieving highly reproducible intermediate $<E_{terminal}>$ in mixed-halide perovskite lighting applications.

Under pulsed laser illumination, observed $\langle E_{\mathrm{PL}}\rangle$ behavior arises from a dynamic equilibrium between forward photosegregation and dark remixing. Observed intermediate $\langle E_{\mathrm{terminal}}\rangle$ dependencies with pulsed excitation repetition rate/duty cycle and peak fluence can be understood within the context of a kinetic model based around a dynamic equilibrium between forward photosegregation and reverse dark remixing. The developed model, in turn, provides semiquantitative estimates of intermediate $\langle E_{\mathrm{terminal}}\rangle$ and corresponding $x_{terminal}$-values and has practical application towards on-demand, color tuning of mixed-halide perovskites. More fundamentally, the established model yields deeper insight into lead-based, mixed-halide perovskite ionic photoinstabilities and may ultimately help explain, yet to be understood phenomena, such as spectral blueshifting under high intensity, pulsed illumination.

**Acknowledgements**
M.K. and P.V.K. thank the Division of Materials Sciences and Engineering, Office of Basic Energy Sciences, U.S. Department of Energy (DOE) under Award DE-SC0014334 for financial support of this work. A.R. thanks the National Energy Research Scientific Computing Center (NERSC) for generous support in the form of computing resources and guidance on high-performance computing for this project. C.D. thanks the Vincent P. Slatt Fellowship for Undergraduate Research in Energy Systems and Processes, which is administered by Notre Dame Energy.

**Competing Interests**
Anthony Ruth is the founder of Chromatic Lighting. Chromatic Lighting is a company dedicated to the commercialization of color-tunable mixed-halide perovskite light emitters. Anthony Ruth has applied for patent protection of color-tunable mixed-halide perovskite light emitters under US patent application number 19/240,949. This patent is pending at the time of publication.

**ORCID**
Anthony Ruth: 0000-0002-2670-5709
Halyna Okrepka: 0000-0002-3165-8521

Michele Vergari: 0000-0001-7465-1445
Charlie Desnoyers: 0000-0001-8285-0081
Minh Nguyen: 0000-0002-9043-9580
Luca Gavioli: 0000-0003-2782-7414
Prashant Kamat: 0000-0002-2465-6819
Masaru Kuno: 0000-0003-4210-8514

[14] Limmer, D. T; Ginsberg, S. N. Photoinduced Phase Separation in the Lead Halides is a Polaronic Effect. *J. Chem. Phys.* **2022**, 152, 230901.
[15] Wang, X.; Ling, Y.; Lian, X.; Xin, Y.; Dhungana, K. B.; Perez-Orive, F.; Knox, J.; Chen, Z.; Zhou, Y.; Beery, D.; Hanson, K.; Shi, J.; Lin, S.; Gao, H. Suppressed Phase Separation of Mixed-Halide Perovskites Confined in Endotaxial Matrices. *Nature Commun.* **2019**, 10, 695.
[16] Knight, A. J; Wright, A. D; Patel J. A; McMeekin, D. P; Snaith H. J; Johnston M. B; Herz L. M. Electronic Traps and Phase Segregation in Lead Mixed-Halide Perovskite. *ACS Energy Lett.* **2019**, 4, 75-84.
[17] Belisle, R. A.; Bush, K. A.; Bertoluzzi, L.; Gold-Parker, A.; Toney, M. F.; McGehee, M. D. Impact of Surfaces on Photoinduced Halide Segregation in Mixed-Halide Perovskites. *ACS Energy Lett.* **2018**, 3, 2694-2700.
[18] Barker, A. J.; Sadhanala, A.; Deschler, F.; Gandini, M.; Senanayak, S. P.; Pearce, P. M.; Mosconi, E.; Pearson, A. J.; Wu, Y.; Kandada, A. R. S.; Leijtens, T.; De Angelis, F.; Dutton, S. E.; Petrozza, A.; Friend, R. H. Defect-Assisted Photoinduced Halide Segregation in Mixed-Halide Perovskite Thin Films. *ACS Energy Lett.* **2017**, 2, 1416-1424.
[19] Pavlovetc, I. M; Ruth, A; Gushchina, I; Ngo, L; Zhang, S; Zhang, Z; Kuno, M. Distinguishing Models for Mixed Halide Lead Perovskite Photosegregation via Terminal Halide Stoichiometry. *ACS Energy Lett.* **2021**, 6, 2064-2071.
[20] Yoon, S. J.; Draguta, S.; Manser, J. S.; Sharia, O.; Schneider, W. F.; Kuno, M.; Kamat, P. V. Tracking Iodide and Bromide Ion Segregation in Mixed Halide Lead Perovskites during Photoirradiation. *ACS Energy Lett.* **2016**, 1, 290-296.
[21] Yoon, S. J.; Kuno, M.; Kamat, P. V. Shift Happens. How Halide Ion Defects Influence Photoinduced Segregation in Mixed Halide Perovskites. *ACS Energy Lett.* **2017**, 2, 1507-1514.
[22] Elmelund, T.; Seger, B.; Kuno, M. K.; Kamat, P. V. How Interplay between Photo and Thermal Activation Dictates Halide Ion Segregation in Mixed Halide Perovskites. *ACS Energy Lett.* **2020**, 5, 56-63.
[23] Mao, W.; Hall, C.; Bernardi, S.; Cheng, Y.; Widmer-Cooper, A.; Smith, T.; Bach, U. Light-Induced Reversal of Ion Segregation in Mixed-Halid Perovskites. *Nature Mater*. **2021**, 20, 55-61.
[24] Okrepka, H; Franzini, L; Pagliara, S; Kuno, M. Photoremixing of Photosegregated Formamidinium/Cesium Lead Iodide/Bromide Thin Films under Pulsed Laser Excitation. *ACS Energy Lett.* **2024**, 9, 5744-5746.
[25] Brennan, M. C.; Ruth, A.; Kamat, P. V.; Kuno, M. Photoinduced Anion Segregation in Mixed Halide Perovskites. *Trends in Chemistry* **2020**, 2, 282-301.
[26] Knight, A.; Borchert, J.; Oliver, R.; Patel, J.; Radaelli, P.; Snaith, H.; Johnston, M.; Herz, L. M. Halide Segregation in Mixed-Halide Perovskites: Influence of A-site Cations. *ACS Energy Lett.* **2021**, 6, 799-808.
[27] Hu, M.; Bi, C.; Yuan, Y.; Bai, Y.; Huang, J. Stabilized Wide Bandgap $MAPbBr_xI_{3-x}$ Perovskite by Enhanced Grain Size and Improved Crystallinity. *Adv. Sci.* **2016**, 3, 1500301.
[28] Eames, C.; Frost, J. M.; Barnes, P. R. F.; O'Regan, B. C.; Walsh, A.; Islam, M. S. Ionic Transport in Hybrid Lead Iodide Perovskite Solar Cells. *Nat. Commun.* **2015**, 6, 7497.
[29] Mizusaki, J.; Arai, K.; Fueki, K. Ionic Conduction of the Perovskite-Type Halides. *Solid State Ionics* **1983**, 11, 203-211.
[30] Kuku, T. A.; Salau, A. M. Electrical Conductivity of $CuSnI_3$, $CuPbI_3$ and $KPbI_3$. *Solid State Ionics* **1987**, 25, 1-7.